\newlength{\absize}
\documentclass[12pt]{article}
\usepackage{latexsym}
\usepackage{amssymb}
\newcommand{\tr}{\mathop{\rm Tr}\nolimits}
\newcommand{\figsize}{\small}

\input prepictex
\input pictex
\input postpictex
\newdimen\tdim
\tdim=\unitlength
\def\stpltsmbl{\setplotsymbol ({\small .})}

\newbox\sru
\setbox\sru=\hbox{\beginpicture
\setcoordinatesystem units <\tdim,\tdim>
\stpltsmbl
\setquadratic
\plot
  0.0   0.0
  4.8   1.5
  7.5   5.0
  7.3   8.5
  5.0  10.0
  2.7   8.5
  2.5   5.0
  5.2   1.5
 10.0   0.0
/
\endpicture}
\def\springru #1 #2 *#3 /{\multiput {\copy\sru}  at
#1 #2 *#3 10 0 /}
\newcommand{\np}{\mbox{\tiny$N$$+$$1$}}
\usepackage{hyperref}

\begin{document}

\thispagestyle{empty}
\pagestyle{empty}
\renewcommand{\thefootnote}{\fnsymbol{footnote}}
\newcommand{\starttext}{\newpage\normalsize
 \pagestyle{plain}
 \setlength{\baselineskip}{3ex}\par
 \setcounter{footnote}{0}
 \renewcommand{\thefootnote}{\arabic{footnote}}
 }
\newcommand{\preprint}[1]{\begin{flushright}
 \setlength{\baselineskip}{3ex}#1\end{flushright}}
\renewcommand{\title}[1]{\begin{center}\LARGE
 #1\end{center}\par}
\renewcommand{\author}[1]{\vspace{2ex}{\Large\begin{center}
 \setlength{\baselineskip}{3ex}#1\par\end{center}}}
\renewcommand{\thanks}[1]{\footnote{#1}}
\renewcommand{\abstract}[1]{\vspace{2ex}\normalsize\begin{center}
 \centerline{\bf Abstract}\par\vspace{2ex}\parbox{\absize}{#1
 \setlength{\baselineskip}{2.5ex}\par}
 \end{center}}

\title{Fun with Higgsless Theories}
\author{
 Howard~Georgi,\thanks{\noindent \tt georgi@physics.harvard.edu}
 \\ \medskip
 Lyman Laboratory of Physics \\
 Harvard University \\
 Cambridge, MA 02138
 }
\date{04/07}
\abstract{Motivated by recent works on ``Higgsless theories,'' I 
discuss an $SU(2)_0\times SU(2)^{N}\times U(1)$ gauge theory with arbitrary
bifundamental (or custodial $SU(2)$ preserving) symmetry breaking between
the gauge subgroups and with ordinary matter transforming only under the
$U(1)$ and $SU(2)_0$.
When the couplings, $g_j$, of the other $SU(2)$s are very large, this
reproduces the standard model at the tree level. 
I calculate the $W$ and $Z$ masses and other electroweak parameters 
in a perturbative 
expansion in $1/g_j^2$, and give physical interpretations of the results
in a mechanical analog built out of masses and springs. In the mechanical
analog, it is clear that even for arbitrary patterns of symmetry breaking,
it is not possible (in the perturbative regime) to raise the Higgs mass
by a large factor while keeping the $S$ parameter small.} 

\starttext

\setcounter{equation}{0}
\section{\label{intro}Higgsless theories, deconstruction, masses, springs}

So-called ``Higgsless'' theories~\cite{Csaki:2003dt, Csaki:2003zu}
make use of boundary conditions on an extra
dimension to break the
electroweak symmetry of the standard model. In a phenomenologically
successful model of this kind (if such could be constructed), there would
be no light scalars, but instead one would 
find additional massive vector bosons
at the electroweak symmetry breaking scale, the Kaluza-Klein
partners of the $W$ and $Z$ from the extra dimension.

A number of groups (see for example \cite{Foadi:2003xa} and
\cite{Chivukula:2004pk}) have studied Higgsless theories using the
technique of deconstruction~\cite{Arkani-Hamed:2001ca, Hill:2000mu} to
actualize the extra-dimensional metaphor in conventional four dimensional
quantum field theory. These works are the primary motivation for this
note. The approach 
here differs from that of previous works in several ways. I
consider a more general pattern of symmetry breaking, preserving a
custodial $SU(2)$ symmetry~\cite{Sikivie:1980hm}, but otherwise completely
arbitrary.\footnote{Reference \cite{Chivukula:2004pk} generalizes the
simple deconstruction in a different direction, including additional $U(1)$
gauge groups, but still retaining the local structure of symmetry breaking
associated with deconstruction.} I analyze these models in a power series
expansion around a 
standard model limit. This is a strong-coupling expansion in the couplings
of the ``extra'' $SU(2)$ gauge groups. I also make use of what I think is
an interesting trick 
to relate the $W$ and $Z$ properties in this general class of
theories. Finally, I discuss a mechanical analog of the field theories in
systems of masses and springs. I believe that this is
extremely useful in developing intuition about the properties of these
theories. In particular, I find physical interpretations of the two most
critical issues facing theories of this kind: raising the scale of symmetry
breaking and 
keeping
the $S$ parameter small. Sadly, I conclude, in agreement with
previous analyses, that the promise of Higgsless theories is unlikely to be
realizable, even in this more general class of theories. But I hope that
the reader will find that this analysis is sufficiently unusual to justify
the term ``fun'' in the title.

In 
section~\ref{where}, I introduce the class of models I discuss in this
paper and briefly discuss the scalar sector.
In section~\ref{springs}, I introduce the mechanical analog - 
two systems of
masses and springs - one related to the $Z$ mass matrix and 
the other to the $W$.
In sections~\ref{lightw} and
section~\ref{lightz}, I study the $W$ and $Z$ mass matrices,
respectively. The analysis of the light $W$ mass is straightforward in
a strong-coupling expansion of the inverse mass matrix around the standard
model limit. A similar analysis of the $Z$ is possible after a
transformation of the inverse mass matrix.
In section~\ref{alphabet}, I discuss the phenomenology of the class
of models by calculating the electroweak parameters, $S$, $T$ and $U$, of
which $S$ is the potential problem. 
Finally in section~\ref{mechanicals}, I give a physical interpretation of
$S$ in the mechanical analog that makes it obvious that $S$ is a very
strong constraint for all models in the class.

\setcounter{equation}{0}\section{\label{where}Where is the Higgs?}

The class of theories that we consider in this paper are
$SU(2)_0\times SU(2)^{N}\times U(1)_{\mbox{\tiny$N$$+$$1$}}$ gauge theories
with arbitrary 
bifundamental (or custodial $SU(2)$ preserving) symmetry breaking between
the gauge subgroups and with ordinary matter transforming only under the
$U(1)_{\mbox{\tiny$N$$+$$1$}}$ and $SU(2)_0$. This includes the
deconstructed version of 
Higgsless theories,~\cite{Foadi:2003xa, Chivukula:2004pk} the Moose diagram
for which is shown in figure~\ref{fig-1}.
We will refer to
this special case
as the ``linear model'' for reasons that are probably obvious.
{\figsize\begin{figure}[htb]
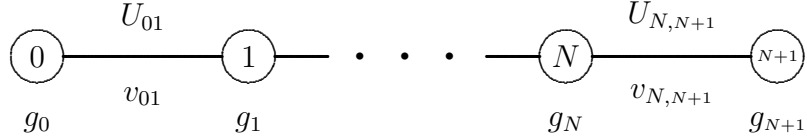

$$\beginpicture
\setcoordinatesystem units <\tdim,\tdim>
\circulararc 360 degrees from 10 0 center at 0 0
\circulararc 360 degrees from 90 0 center at 80 0
\circulararc 360 degrees from 210 0 center at 200 0
\circulararc 360 degrees from 290 0 center at 280 0
\put {$0$} at 0 0
\put {$1$} at 80 0
\put {\tiny$N$$+$$1$} at 280 0
\put {$N$} at 200 0
\stpltsmbl
\plot 10 0 70 0 /
\plot 110 0 90 0 /
\plot 170 0 190 0 /
\plot 210 0 270 0 /
\put {$U_{01}$} at 40 15
\put {$U_{N,\np}$} at 240 15
\put {$v_{01}$} at 40 -15
\put {$v_{N,\np}$} at 240 -15
\put {$g_0$} at 0 -25
\put {$g_1$} at 80 -25
\put {$g_{\np}$} at 280 -25
\put {$g_N$} at 200 -25
\multiput {\tiny$\bullet$} at 122 0 *2 17 0 /
\endpicture$$
\caption{\figsize\sf\label{fig-1}The Moose diagram associated with the linear
model.}\end{figure}} 
More general patterns of symmetry 
breaking can involve additional links between
nodes of the Moose. For example, for $N=1$, there are three other distinct
possibilities. They are shown in figure~\ref{fig-1-3}.
{\figsize\begin{figure}[htb]
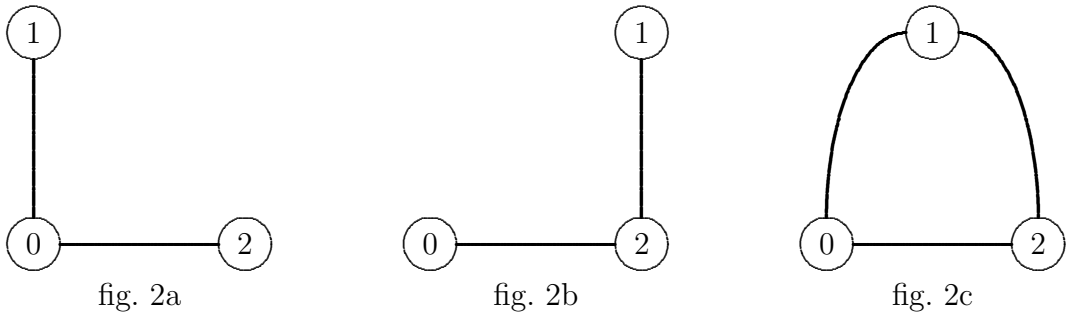

$$\beginpicture
\setcoordinatesystem units <\tdim,\tdim>
\put {\beginpicture
\setcoordinatesystem units <\tdim,\tdim>
\circulararc 360 degrees from 10 0 center at 0 0
\circulararc 360 degrees from 90 0 center at 80 0
\circulararc 360 degrees from 10 80 center at 0 80
\put {$0$} at 0 0
\put {$2$} at 80 0
\put {$1$} at 0 80
\stpltsmbl
\plot 10 0 70 0 /
\plot 0 10 0 70 /
\put {fig.~\protect\ref{fig-1-3}a} at 40 -20
\linethickness=0pt
\putrule from 0 -25 to 0 90
\endpicture} at 0 0
\put {\beginpicture
\setcoordinatesystem units <\tdim,\tdim>
\circulararc 360 degrees from 10 0 center at 0 0
\circulararc 360 degrees from 90 0 center at 80 0
\circulararc 360 degrees from 90 80 center at 80 80
\put {$0$} at 0 0
\put {$2$} at 80 0
\put {$1$} at 80 80
\stpltsmbl
\plot 10 0 70 0 /
\plot 80 10 80 70 /
\put {fig.~\protect\ref{fig-1-3}b} at 40 -20
\linethickness=0pt
\putrule from 0 -25 to 0 90
\endpicture} at 150 0
\put {\beginpicture
\setcoordinatesystem units <\tdim,\tdim>
\circulararc 360 degrees from 10 0 center at 0 0
\circulararc 360 degrees from 90 0 center at 80 0
\circulararc 360 degrees from 50 80 center at 40 80
\put {$0$} at 0 0
\put {$2$} at 80 0
\put {$1$} at 40 80
\stpltsmbl
\plot 10 0 70 0 /
\ellipticalarc axes ratio 1:2.33 90 degrees from 30 80 center at 30 10 
\ellipticalarc axes ratio 1:2.33 -90 degrees from 50 80 center at 50 10 
\put {fig.~\protect\ref{fig-1-3}c} at 40 -20
\linethickness=0pt
\putrule from 0 -25 to 0 90
\endpicture} at 300 0
\endpicture$$
\caption{\figsize\sf\label{fig-1-3}The three ``non-linear'' (meaning not
equivalent to the linear model) symmetry breaking patterns for
$N=1$.}\end{figure}} 
The number of possible symmetry breaking patterns grows very rapidly with
$N$. For $N=2$, there are fifteen, one of which is shown in
figure~\ref{fig-1-4}. 
{\figsize\begin{figure}[htb]
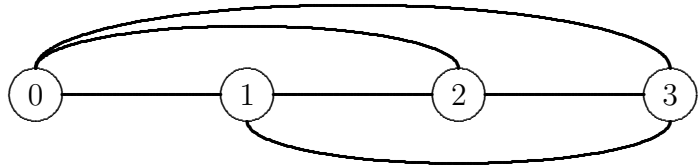

$$\beginpicture
\setcoordinatesystem units <\tdim,\tdim>
\circulararc 360 degrees from 10 0 center at 0 0
\circulararc 360 degrees from 90 0 center at 80 0
\circulararc 360 degrees from 170 0 center at 160 0
\circulararc 360 degrees from 250 0 center at 240 0
\put {$0$} at 0 0
\put {$1$} at 80 0
\put {$2$} at 160 0
\put {$3$} at 240 0
\stpltsmbl
\plot 10 0 70 0 /
\plot 90 0 150 0 /
\plot 170 0 230 0 /
\ellipticalarc axes ratio 5:1 180 degrees from 240 10 center at 120 10
\ellipticalarc axes ratio 5:1 180 degrees from 160 10 center at 80 10
\ellipticalarc axes ratio 5:1 180 degrees from 80 -10 center at 160 -10
\linethickness=0pt
\putrule from 0 -30 to 0 70
\endpicture$$
\caption{\figsize\sf\label{fig-1-4}A Moose diagram for general symmetry
breaking for $N=2$. There are fourteen other distinct possibilities
(counting the linear model) in which some of the links are
missing.}\end{figure}}
In this paper, I am not interested in doing away with the Higgs
entirely. I am happy to think about the symmetry breaking being done by the
vacuum expectation values (VEVs) of scalar fields.
The question I address is whether we can raise the lightest
scalar mass above the TeV scale while
retaining the phenomenology of the standard model. Since this would give
rise to a Higgsless effective low energy theory, we will continue to use the
term Higgsless.

These theories 
involve several independent symmetry breaking
sectors. This raises the question, given a set of symmetry breaking
sectors (SBSs), 
where do we expect the lightest scalar ``Higgs''? 
Let's briefly consider this in the simple
realization in which each symmetry breaking sector is just a linear
$\sigma$-model. Here there is a single 
neutral custodial $SU(2)$ singlet scalar for each $\sigma$-model, and
depending on the structure of the theory, there may be custodial $SU(2)$
triplet scalar pseudo-Goldstones below the cut-off scale. 
The obvious thing to say, I
think, is that if we have a set of SBSs with scales $v_j$,
the lightest scalar would be expected at or below the lowest
symmetry breaking scale
\begin{equation}
\approx4\pi\min_j v_j
\end{equation}

If this were just an ordinary field theory with several $\sigma$-models,
there would be arbitrary couplings without respect to locality. Then all
the VEVs would be of the same order of magnitude unless some fine tuning
was going on. But in an extra dimension 
stretched ``between'' the $SU(2)_0$ and the $U(1)_{\mbox{\tiny$N$$+$$1$}}$,
locality is a strong constraint. 
First of all, with locality, we don't have to worry about
pseudo-Goldstones. They are all eaten.
Also because of locality, we can imagine
some dependence of the VEV on ``position'' in the extra
dimension. However, in this case, it might be argued that the Higgs would
show up at the smallest scale, so it probably makes sense to keep all the
scales the same if we are trying to push up the Higgs mass as much as
possible. 

Without locality, the situation is more complicated, but it does not seem
to be any {\bf better}, at least not if the goal is to push up the 
minimum mass of things in the scalar
sector. We will ignore this and make the simple assumption
that all the symmetry breaking scales are of the same order of magnitude.

\setcounter{equation}{0}\section{\label{springs}Springs and masses}

There is a mechanical analog to each of the theories we consider.
We can think of each gauge group as a degree of freedom with mass $1/g^2$
and each VEV between groups as a 
massless spring with spring constant $v^2$. Then the
masses of the gauge bosons are proportional 
to the frequencies of the normal modes.

To see how this works more precisely, let's look at 
the example of the linear model illustrated in
figure~\ref{fig-1}, where the groups associated with nodes $0$-$N$ are
$SU(2)$s and the group associated with $N+1$ is a
$U(1)_{\mbox{\tiny$N$$+$$1$}}$.\footnote{As much 
as possible, I use the notation of reference \cite{Chivukula:2004pk}, 
though I will discuss only the case of a single $U(1)$.} 
The mechanical analog of the neutral gauge boson sector is then illustrated
in figure~\ref{fig-2}. 
{\figsize\begin{figure}[htb]
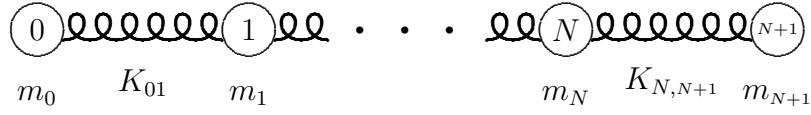

$$\beginpicture
\setcoordinatesystem units <\tdim,\tdim>
\circulararc 360 degrees from 10 0 center at 0 0
\circulararc 360 degrees from 90 0 center at 80 0
\circulararc 360 degrees from 210 0 center at 200 0
\circulararc 360 degrees from 290 0 center at 280 0
\put {$0$} at 0 0
\put {$1$} at 80 0
\put {\tiny$N$$+$$1$} at 280 0
\put {$N$} at 200 0
\stpltsmbl
\springru 10 -4 *5 /
\springru 90 -4 *1 /
\springru 170 -4 *1 /
\springru 210 -4 *5 /
\put {$K_{01}$} at 40 -20
\put {$K_{N,\np}$} at 240 -20
\put {$m_0$} at 0 -25
\put {$m_1$} at 80 -25
\put {$m_N$} at 200 -25
\put {$m_{\np}$} at 280 -25
\multiput {\tiny$\bullet$} at 122 0 *2 17 0 /
\endpicture$$
\caption{\figsize\sf\label{fig-2}The mechanical analog of the system in
figure~\protect\ref{fig-1}.}\end{figure}} 

In the linear model, we can write the neutral gauge boson mass-squared
matrix as the $N$$+$$2\times N$$+$$2$ matrix
\begin{equation}
M_n^2=\frac{1}{4}\,G\,V\,G
\label{mn2}
\end{equation}
where $G$ is the diagonal matrix of gauge couplings
\begin{equation}
G=\pmatrix{
g_0&0&\cdots&0&0\cr
0&g_1&\cdots&0&0\cr
\vdots&\vdots&\ddots&\vdots&\vdots\cr
0&0&\cdots&g_N&0\cr
0&0&\cdots&0&g_{\np}\cr
}
\label{g}
\end{equation}
and the matrix $V$ is
\begin{equation}
V=\pmatrix{
v^2_{01}&-v^2_{01}&\cdots&0&0\cr
-v^2_{01}&v^2_{01}+v^2_{12}&\cdots&0&0\cr
\vdots&\vdots&\ddots&\vdots&\vdots\cr
0&0&\cdots&
v^2_{\mbox{\tiny$N$$-$$1$},N}
+v^2_{N,\np}&-v^2_{N,\np}\cr
0&0&\cdots&-v^2_{N,\np}&v^2_{N,\np}\cr
}
\label{v}
\end{equation}
These VEV's break the gauge symmetry down to a single diagonal symmetry if
all the gauge groups are the same. 
The squared gauge boson masses,
$m^2_\alpha$ and the 
corresponding mass eigenstates, $\kappa^\alpha$ are eigenvalues and
eigenvectors of the gauge boson mass squared
matrix, 
\begin{equation}
\frac{1}{4}G\,V\,G\,\kappa^\alpha=m_\alpha^2\,\kappa^\alpha
\label{gvg}
\end{equation}

For more general symmetry breaking, the formulas are the same (so long as
each symmetry breaking sector preserves a custodial $SU(2)$ and
``plaquette'' terms are introduced to align the vacuum properly), 
except that
more entries in the VEV matrix $V$ are populated. I will analyze the
general case, but will continue also to illustrate the analysis in the
simple example of a linear theory space.

For the mechanical analog, the squared normal angular frequencies,
$\omega^2_\alpha$ and the 
corresponding normal
modes, $\lambda^\alpha$ are eigenvalues and eigenvectors of the $M^{-1}K$
matrix, 
\begin{equation}
M^{-1}K\,\lambda^\alpha=\omega_\alpha^2\,\lambda^\alpha
\label{m-1k}
\end{equation}
where
\begin{equation}
M=\pmatrix{
m_0&0&\cdots&0&0\cr
0&m_1&\cdots&0&0\cr
\vdots&\vdots&\ddots&\vdots&\vdots\cr
0&0&\cdots&m_N&0\cr
0&0&\cdots&0&m_{\np}\cr
}
\label{m}
\end{equation}
and the matrix $K$ is
\begin{equation}
K=\pmatrix{
K_{01}&-K_{01}&\cdots&0&0\cr
-K_{01}&K_{01}+K_{12}&\cdots&0&0\cr
\vdots&\vdots&\ddots&\vdots&\vdots\cr
0&0&\cdots&
K_{\mbox{\tiny$N$$-$$1$},N}
+K_{N,\np}&-K_{N,\np}\cr
0&0&\cdots&-K_{N,\np}&K_{N,\np}\cr
}
\label{k}
\end{equation}

We can rewrite (\ref{gvg}) as
\begin{equation}
G^2\,V\,(G\,\kappa^\alpha)=m_\alpha^2\,(G\,\kappa^\alpha)
\label{gvg2}
\end{equation}
and in this form it is clear that there is an exact correspondance,
\begin{equation}
\frac{1}{4}G^2
\leftrightarrow
M^{-1}
\quad\quad
V
\leftrightarrow
K
\quad\quad
G\,\kappa^\alpha
\leftrightarrow
\lambda^\alpha
\quad\quad
m_\alpha
\leftrightarrow
\omega_\alpha
\label{analog}
\end{equation}

Notice that if the mass of a degree of freedom is very small and if
there are only two springs attached, it is as if
there is a single, continuous spring with no mass on it between the degrees
of freedom on either side. The spring constants then add reciprocally, like
capacitances.
For example, suppose $g_2$ goes to infinity.
The effective
spring constant between the two nodes $1$ and $3$ is then
\begin{equation}
\frac{1}{1/v_{12}^2+1/v_{23}^2}
\label{likek}
\end{equation}

At the level of the Goldstone bosons, the relevant Goldstone kinetic energy
term are
\begin{equation}
\begin{array}{c}
\displaystyle
\frac{v_{12}^2}{4}\tr\left([\partial^\mu U_{12} -igU_{12}W^\mu]
[\partial_\mu U_{12} -igU_{12}W_\mu]^\dagger\right)
\\ \displaystyle
+\frac{v_{23}^2}{4}\tr\left([\partial^\mu U_{23} +igW^\mu U_{23}]
[\partial_\mu U_{23} +igW_\mu U_{23}]^\dagger\right)
\end{array}
\end{equation}
The eaten Goldstone boson is
\begin{equation}
\frac{-i}{2}\left(v_{12}^2\,[\partial^\mu U_{12}]^\dagger U_{12}
-v_{23}^2\,U_{23}[\partial^\mu U_{23}]^\dagger\right)
\end{equation}
And the Goldstone boson kinetic energy can be written as
\begin{equation}
\begin{array}{c}
\displaystyle
\frac{1}{4}\frac{1}{1/v_{12}^2+1/v_{23}^2}
\tr\left(
[\partial^\mu(U_{12}U_{23})]
[\partial_\mu(U_{12}U_{23})]^\dagger
\right)
+\frac{1}{4}\frac{1}{v_{12}^2+v_{23}^2}
\\ \displaystyle
\tr\left(
\left(v_{12}^2\,[\partial^\mu U_{12}]^\dagger U_{12}
-v_{23}^2\,U_{23}[\partial^\mu U_{23}]^\dagger\right)
\left(v_{12}^2\,[\partial^\mu U_{12}]^\dagger U_{12}
-v_{23}^2\,U_{23}[\partial^\mu U_{23}]^\dagger\right)^\dagger
\right)
\end{array}
\end{equation}
and you can see that the spring constant of the uneaten Goldstone boson is
given by (\ref{likek}).

If a gauge coupling goes to zero, which is equivalent to having no gauge
symmetry at all, this corresponds to an infinite
mass, which is like a fixed wall. So for example, in the linear model,
the mechanical analog for the neutral gauge bosons looks like
figure~\ref{fig-2}, but for the charged gauge bosons, the mechanical is
shown in figure~\ref{fig-3}. 
{\figsize\begin{figure}[htb]
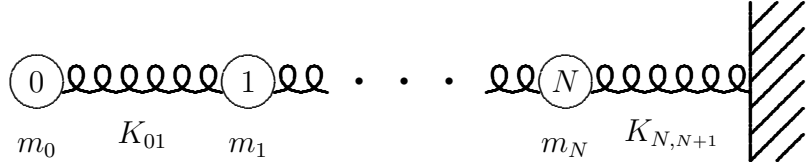

$$\beginpicture
\setcoordinatesystem units <\tdim,\tdim>
\circulararc 360 degrees from 10 0 center at 0 0
\circulararc 360 degrees from 90 0 center at 80 0
\circulararc 360 degrees from 210 0 center at 200 0
\put {$0$} at 0 0
\put {$1$} at 80 0
\put {$N$} at 200 0
\stpltsmbl
\springru 10 -4 *5 /
\springru 90 -4 *1 /
\springru 170 -4 *1 /
\springru 210 -4 *5 /
\put {$K_{01}$} at 40 -20
\put {$K_{N,\np}$} at 240 -20
\put {$m_0$} at 0 -25
\put {$m_1$} at 80 -25
\put {$m_N$} at 200 -25
\multiput {\tiny$\bullet$} at 122 0 *2 17 0 /
\plot 270 -30 270 30 /
\multiput {\beginpicture
\setcoordinatesystem units <\tdim,\tdim>
\plot -10 -10 10 10 /
\endpicture} at 280 -20 *4 0 10 /
\plot 280 -30 290 -20 /
\plot 270 20 280 30 /
\endpicture$$
\caption{\figsize\sf\label{fig-3}The mechanical analog for the charged
gauge boson mass matrix.}\end{figure}}
The more general case would have every spring connected to the $N$$+$$1$st
mass in the $Z$ analog connected to the fixed wall in the $W$ analog.
The corresponding gauge boson mass
squared matrix
for the charged gauge bosons is the $N$$+$$1\times N$$+$$1$ matrix,
obtained from (\ref{mn2}) by eliminating the $N$$+$$2$nd
row and column,
\begin{equation}
M_c^2=\frac{1}{4}\widetilde G\,\widetilde V\,\widetilde G
\label{mc2}
\end{equation}
where $\widetilde G$ is the diagonal matrix of gauge couplings without
$g_{\np}$
\begin{equation}
\widetilde G=\pmatrix{
g_0&0&\cdots&0&0\cr
0&g_1&\cdots&0&0\cr
\vdots&\vdots&\ddots&\vdots&\vdots\cr
0&0&\cdots&g_{\mbox{\tiny$N$$-$$1$}}&0\cr
0&0&\cdots&0&g_N\cr
}
\label{tildeg}
\end{equation}
and the matrix $\widetilde V$ in the linear model is
\begin{equation}
\widetilde V=\pmatrix{
v^2_{01}&-v^2_{01}&\cdots&0&0\cr
-v^2_{01}&v^2_{01}+v^2_{12}&\cdots&0&0\cr
\vdots&\vdots&\ddots&\vdots&\vdots\cr
0&0&\cdots
&v^2_{\mbox{\tiny$N$$-$$2$},\mbox{\tiny$N$$-$$1$}}+
v^2_{\mbox{\tiny$N$$-$$1$},N}
&-v^2_{\mbox{\tiny$N$$-$$1$},N}\cr
0&0&\cdots&-v^2_{\mbox{\tiny$N$$-$$1$},N}
&v^2_{\mbox{\tiny$N$$-$$1$},N}+
v^2_{N,\np}\cr
}
\label{tildev}
\end{equation}
with corresponding $\widetilde M$ and $\widetilde K$ for the mechanical
analog of 
figure~\ref{fig-3}. 

Again, the general formula is analogous. $\tilde V$ is obtained 
from $V$ in
the same way, by removing the $N$$+$$2$nd row and column.

The low energy charged-current
weak interactions are determined by the inverse of $\widetilde
V$,\footnote{Because of our numbering of the
gauge groups, to agree with reference \cite{Chivukula:2004pk}, 
it is convenient to label
the rows and columns of our vectors and matrices 
beginning with $0$ rather than $1$, so that is what
we will do.}
\begin{equation}
\sqrt{2}\, G_F=\frac{1}{v^2}
=[\widetilde V^{-1}]_{00}
\leadsto\sum_{j=0}^{N}\frac{1}{v_{j,j+1}^2}
\label{vi}
\end{equation}
where I have indicated the value of $[\widetilde V^{-1}]_{00}$ in the
linear example by the symbol $\leadsto$. I will continue to use this 
notation below.

The low energy neutral-current weak interactions are then given in terms of
$\widetilde V^{-1}$ by the Georgi-Weinberg
construction~\cite{Georgi:1977wk} 
(assuming that matter
couples only to $0$ and $N+1$)
\begin{equation}
\sum_{j,k=0}^{N}[\widetilde V^{-1}]_{jk}
\left[T_3\delta_{j0}-\frac{e^2}{g_j^2}Q\right]\,
\left[T_3\delta_{k0}-\frac{e^2}{g_k^2}Q\right]
\label{nc}
\end{equation}
For convenience in the following, I will sometimes abbreviate
the matrix elements $[\widetilde V^{-1}]_{jk}$ as follows:
\begin{equation}
[\widetilde V^{-1}]_{00}=\frac{1}{v^2}\equiv \chi_0\,,
\quad
[\widetilde V^{-1}]_{j0}=
[\widetilde V^{-1}]_{0j}\equiv \chi_j\,,
\quad
[\widetilde V^{-1}]_{jk}\equiv \chi_{jk}
\quad\mbox{for $j,k=1$ to $N$}
\label{chi-gen}
\end{equation}
In the linear theory,
\begin{equation}
\chi_j\leadsto\sum_{\ell=j}^N\frac{1}{v_{\ell,\ell+1}^2}
\quad\mbox{and}\quad
\chi_{jk}\leadsto\chi_{\max(j,k)}
\label{chi}
\end{equation}

The premise of Higgsless models (in their deconstructed form)
is that by extending the gauge group to
include additional copies of $SU(2)$ we can raise the scale of all the
symmetry-breaking breaking physics above a TeV, 
thus pushing the Higgs boson out of the
low energy theory, while leaving the $W$ and $Z$ mass and the low energy
weak interactions unchanged. In such a model, the job of unitarizing
$W$-$W$ scattering at a TeV would be done by the extra massive vector
bosons, some of which would necessarily appear below the TeV scale. 

The mechanical analog of the raising of the symmetry-breaking scale
is the
following. You have only very stiff springs (corresponding to a
high symmetry breaking scale), and you want to build a system that has
low frequency normal modes (the $W$ and $Z$) with the same properties as 
those in a
system with a single more flexible spring!
It easy to see how we can do this, at least
classically. The linear model works very well for this purpose.
If we string stiff springs together in series with light or massless
connections, the result behaves for low frequencies like a single flexible
spring. Thus
if we could make the gauge couplings
$g_1$-$g_N$ very large, we could
break up the spring into segments, each of which has a larger spring
constant and therefore larger Higgs mass. In the limit
\begin{equation}
g_j\to\infty\quad\mbox{for $j=1$ to $N$}
\end{equation}
only $[\widetilde V^{-1}]_{00}$ is relevant to the low energy weak
interactions. This is a deconstructed version the strong coupling limit of
a Higgsless model.\footnote{See for example \cite{Foadi:2003xa}.} 

\setcounter{equation}{0}\section{\label{lightw}The light $W$ mass}

We can find the light $W$ 
gauge boson mass by diagonalizing
the inverse mass squared matrix,
\begin{equation}
4\widetilde G^{-1}\,\widetilde V^{-1}\,\widetilde G^{-1}
\approx
\pmatrix{
4\chi_0/g_0^2&0&\cdots\cr
0&0&\cdots\cr
\vdots&\vdots&\ddots\cr
}
=\pmatrix{
4/g_0^2v^2&0&\cdots\cr
0&0&\cdots\cr
\vdots&\vdots&\ddots\cr
}
\label{mw2inv}
\end{equation}
which in the linear model 
depends only on the sum of the reciprocal VEVs (see
(\ref{vi})). The advantage of working with the inverse mass 
squared matrix rather
than the mass squared matrix itself is that in the limit we are
considering, the inverse light $W$ mass squared dominates the inverse
matrix which as you see in (\ref{mw2inv}), is automatically diagonal in the
limit.

Somewhat less naively, we should not allow the other couplings to be
infinitely large. Presumably the picture ceases to make sense if
the $g_j$ are
larger than or of order $4\pi$. 
This means that we cannot take our connectors in the mechanical analog to
be massless. They have some minimum possible mass (not such an unreasonable
toy model).

In the inverse mass squared matrix,
we can easily include 
the effects of the other
couplings to second order using 
ordinary perturbation theory for the inverse mass squared 
matrix. Because the
largest eigenvalue is non-degenerate, 
we can immediately write down the corrections for this state. To second
order in $g_0/g_j$ (for $j=1$ to $N$),
the $W$ eigenvector $\tilde\kappa^0$ is
approximately given by
\begin{equation}
[\tilde\kappa^0]_j\propto
[\widetilde G^{-1}\tilde\lambda^0]_j
\propto[\widetilde K^{-1}]_{j0}/g_j=
\frac{\chi_j}{g_j}\leadsto
\frac{1}{g_j}\sum_{k=j}^{N}\frac{1}{v_{k,k+1}^2}
\label{weigenvector}
\end{equation}
and the eigenvalue is
\begin{equation}
\frac{1}{M_W^2}
=\frac{4\chi_0}{g_0^2}
+\frac{4}{\chi_0}\,\sum_{j=1}^{N}\frac{\chi_j^2}{g_j^2}
=\frac{4}{\chi_0}\,\sum_{j=0}^{N}\frac{\chi_j^2}{g_j^2}
=4v^2\,\sum_{j=0}^{N}\frac{\chi_j^2}{g_j^2}
\label{mw2i}
\end{equation}
It is a little curious that in this expression, the large $1/g_0^2$ term
just seems to be one of a series of terms with the same structure. 

There is a simple physical argument for (\ref{weigenvector}) based on
the mechanical analog (you can refer to figure~\ref{fig-3} to see how this
work in the linear model, but remember that the discussion works for the
general case).
It is clear
that the low frequency mode in the limit in which
$m_0$ is much bigger than all the other masses is approximately just a
static stretching of the springs, with no force on any of the masses except
$0$. Thus this mode satisfies
\begin{equation}
F_j\propto [\widetilde K \,\tilde\lambda]_j\propto\delta_{j0}
\end{equation}
and thus
\begin{equation}
[\tilde\lambda^0]_j\propto [\widetilde K^{-1}]_{j0}=
\chi_j
\label{stretch}
\end{equation}
The dictionary (\ref{analog}) then immediately implies (\ref{weigenvector}).
The mass (\ref{mw2i}) is the expectation value of the inverse
mass-squared matrix in the state (\ref{weigenvector}).

The heavy states are initially degenerate, and to second order the $N\times
N$ inverse mass squared matrix has matrix elements
\begin{equation}
4\,\frac{1}{g_j}\,\left(
\chi_{jk}-\frac{\chi_j\,\chi_k}{\chi_0}
\right)\,\frac{1}{g_k}
\quad\mbox{for $j,k=1$ to $N$.}
\label{heavy-w-matrix}
\end{equation}
An interesting quantity that I will discuss later is the sum of the
inverse mass squares, given by the trace. This simplifies in the linear model:
\begin{equation}
\sum_{{\rm heavy}\atop W{\rm s} }\frac{1}{4M^2}
=
\sum_{j=0}^N\frac{\chi_{jj}}{g_j^2}
-\frac{1}{\chi_0}\sum_{j=0}^N\frac{\chi_j^2}{g_j^2}
\leadsto
\sum_{j=0}^N\frac{\chi_j}{g_j^2}
-\frac{1}{\chi_0}\sum_{j=0}^N\frac{\chi_j^2}{g_j^2}
\label{hw}
\end{equation}

\setcounter{equation}{0}\section{\label{lightz}The light $Z$ mass}

Now we need to find the light 
$Z$ mass. The neutral mass squared matrix given by (\ref{mn2}) has, of
course, a zero eigenvalue associated with the photon. 
The photon eigenstate, as usual, is
\begin{equation}
\kappa^{\np}=\pmatrix{
e/g_0\cr
\vdots\cr
e/g_j\cr
\vdots\cr
e/g_{\np}\cr
}
\quad\quad
\lambda^{\np}\propto\pmatrix{
1\cr
\vdots\cr
1\cr
\vdots\cr
1\cr
}
\end{equation}
Again, it is easiest to work with the inverse mass squares.
The neutral gauge boson mass-squared matrix is not invertible because of
the photon, but we can invert it on the subspace orthogonal to the photon
eigenvector, $\kappa^{\np}$,
and this can be written in terms of $\widetilde V^{-1}$. This is the basis of
the Georgi-Weinberg construction.~\cite{Georgi:1977wk}

Define $\widetilde{\widetilde V}$ and $\widetilde{\widetilde
V}^{\mbox{\scriptsize``$-1$''}}$ as\footnote{I have put the superscript 
$-1$ in
quotes to indicate that is only the inverse on the subspace
orthogonal to the $N+1$ direction.} 
\begin{equation}
\left[\widetilde{\widetilde V}\right]_{jk}
=\left[{\widetilde V}\right]_{jk}\quad\mbox{for $j,k=0$ to $N$}
\quad\mbox{and}\quad 
\left[\widetilde{\widetilde V}\right]_{j,\np}
=\left[\widetilde{\widetilde V}\right]_{\np,k}
=0
\label{ttv}
\end{equation}
\begin{equation}
\left[\widetilde{\widetilde V}^{\mbox{\scriptsize``$-1$''}}\right]_{jk}
=\left[{\widetilde V}^{-1}\right]_{jk}\quad\mbox{for $j,k=0$ to $N$}
\quad\mbox{and}\quad 
\left[\widetilde{\widetilde V}^{\mbox{\scriptsize``$-1$''}}\right]_{j,\np}
=\left[\widetilde{\widetilde V}^{\mbox{\scriptsize``$-1$''}}\right]_{\np,k}
=0
\label{ttvi}
\end{equation}
Then the inverse of the 
neutral gauge boson mass-squared matrix 
on the subspace orthogonal to the photon
eigenvector is
\begin{equation}
{M_n^2}^{\mbox{\scriptsize``$-1$''}}
=4\,(I-\kappa^{\np}{\kappa^{\np}}^T)\,
G^{-1}\,\widetilde{\widetilde V}^{\mbox{\scriptsize``$-1$''}}\,G^{-1}\,
(I-\kappa^{\np}{\kappa^{\np}}^T)
\label{mgw1}
\end{equation}
This can also be written as
\begin{equation}
4\,G^{-1}\,(I-\lambda^{\np}
{\lambda^{\np}}^T\,e^2
G^{-2})\,\widetilde{\widetilde V}^{\mbox{\scriptsize``$-1$''}}\,
(I-e^2G^{-2}\,\lambda^{\np}
{\lambda^{\np}}^T)\,G^{-1}
\label{mgw2}
\end{equation}

In this basis, (\ref{mgw2}) is not diagonal as $g_j\to\infty$.
We need to diagonalize before we
apply perturbation theory. 
But there is a slightly peculiar trick
that allows us to do this automatically,
changing to a more convenient basis without making a mess
of the matrix. I will first describe the trick in general, and then apply
it to (\ref{mgw2}).

Suppose $\kappa$ and ${\hat e}$ are unit vectors. Then
\begin{equation}
P_\pm=\frac{1}{2\left(1\pm \kappa^T {\hat e}\right)}
\,\left(\kappa\pm {\hat e}\right)
\,\left(\kappa\pm {\hat e}\right)^T
\end{equation}
are projection operators onto the one-dimensional subspaces spanned by
$\kappa\pm {\hat e}$ respectively. Then
\begin{equation}
U_\pm=I-\frac{1}{\left(1\pm \kappa^T {\hat e}\right)}
\,\left(\kappa\pm {\hat e}\right)
\,\left(\kappa\pm {\hat e}\right)^T
\label{u}
\end{equation}
are symmetric
orthogonal matrices, with eigenvalue $-1$ on the one-dimensional subspace
and $1$ elsewhere, so they satisfy
\begin{equation}
U_\pm
=U_\pm^T
=U_\pm^{-1}
\label{toggle}
\end{equation}
Since these give opposite signs on $\kappa+{\hat e}$ and
$\kappa-{\hat e}$, they just interchange $\kappa$ and ${\hat e}$. One finds
\begin{equation}
U_\pm\,\kappa=\mp {\hat e}\;,
\quad
U_\pm\,{\hat e}=\mp \kappa
\quad\mbox{and transposes.}
\end{equation}

We are interested is applying this transformation 
to objects which are annihilated by
$\kappa$.
We want to make use of the fact that the components
of such an object in the $e$ direction can always be eliminated in terms of
the other components. We can write
\begin{equation}
\left(I-\kappa\,\kappa^T\right)
=\left(I-\kappa\,\kappa^T\right)
\left[I
-\kappa\,{\hat e}^T/({\hat e}^T\kappa)\,\right]
\end{equation}
because the second term in square brackets vanishes, and
\begin{equation}
=\left(I-\kappa\,\kappa^T\right)
\left(I-{\hat e}\,{\hat e}^T\right)
\left[I
-\kappa\,{\hat e}^T/({\hat e}^T\kappa)\,\right]
\end{equation}
because the second term in the middle factor vanishes.

So now we look at
\begin{equation}
\left(I-\kappa\,\kappa^T\right)\,U_\pm
=\left(I-\kappa\,\kappa^T\right)\,
\left(I-{\hat e}\,{\hat e}^T\right)
\,\left[I
-\kappa\,{\hat e}^T/({\hat e}^T\kappa)\,\right]
\,
U_\pm
\,
\left(I-{\hat e}\,{\hat e}^T\right)
\end{equation}
\begin{equation}
=\left(I-\kappa\,\kappa^T\right)\,
\left(I-{\hat e}\,{\hat e}^T\right)
\,\left[I
-\kappa\,{\hat e}^T/({\hat e}^T\kappa)\,\right]
\,
\left(I\mp \frac{{\hat e}\,\kappa^T}{(1\pm\kappa^T{\hat e})}\right)
\,
\left(I-{\hat e}\,{\hat e}^T\right)
\end{equation}
\begin{equation}
=\left(I-\kappa\,\kappa^T\right)\,
\left(I-{\hat e}\,{\hat e}^T\right)
\,
\left(I\pm 
\frac{\kappa\,\kappa^T}{({\hat e}^T\kappa)\,(1\pm\kappa^T{\hat e})}\right)
\,
\left(I-{\hat e}\,{\hat e}^T\right)
\label{pipm1}
\end{equation}
We can't average these because $U_+$ and $U_-$ are not equal, though
both have similar properties.

We can write (\ref{pipm1}) equivalently as
\begin{equation}
\left(I-\kappa\,\kappa^T\right)\,U_\pm
=\left(I-\kappa\,\kappa^T\right)\,H_\pm
\label{pipm2}
\end{equation}
where
\begin{equation}
H_\pm\equiv \left(
\left(I-{\hat e}\,{\hat e}^T\right)
\pm \frac{\left(\kappa-{\hat e}\,(\kappa^T{\hat e})\,\right)\,
\left(\kappa^T-(\kappa^T{\hat e})\,{\hat e}^T\right)}
{({\hat e}^T\kappa)\,(1\pm\kappa^T{\hat e})}\right)
\label{hpm}
\end{equation}
where we don't need the projectors onto the subspace orthogonal to $\hat e$.

We are interested in the case where $\kappa=\kappa_{\np}$, the photon
eigenvector, and ${\hat e}={\hat e}_{\np}$, the 
unit vector in the $N+1$ direction. Before proceeding, let's check the
result for $N=0$. This is also the $0$th contribution to the $Z$ mass in
the general theory, so we have to do it anyway. In this case, 
\begin{equation}
\kappa=
\pmatrix{
\sin\theta\cr
\cos\theta\cr
}
\quad\mbox{and}\quad
\hat e=\pmatrix{
0\cr
1\cr
}
\end{equation}
Then (\ref{hpm}) becomes
\begin{equation}
H_\pm=\pmatrix{
1&0\cr
0&0\cr
}
\pm\frac{1}{\cos\theta\,(1\pm\cos\theta)}
\pmatrix{
\sin\theta\cr
0\cr
}
\pmatrix{
\sin\theta&
0\cr
}
\end{equation}
\begin{equation}
=\pmatrix{
1&0\cr
0&0\cr
}
\pm\frac{1}{\cos\theta\,(1\pm\cos\theta)}
\pmatrix{
\sin^2\theta&
0\cr
0&0\cr
}
\end{equation}
\begin{equation}
=\pmatrix{
1&0\cr
0&0\cr
}
\pm\frac{1}{\cos\theta\,(1\pm\cos\theta)}
\pmatrix{
1-\cos^2\theta&
0\cr
0&0\cr
}
\end{equation}
\begin{equation}
=\pmatrix{
1&0\cr
0&0\cr
}
\pm\frac{1}{\cos\theta}
\pmatrix{
1\mp\cos\theta&
0\cr
0&0\cr
}
=
\pmatrix{
\pm1/\cos\theta&
0\cr
0&0\cr
}
\end{equation}
There is one such factor from each side of the mass matrix, so this gives the
usual factor of $1/\cos^2\theta$ in the $Z$ mass compared to the $W$
mass.\footnote{Notice that in the mechanical analog for $N=0$, the
$\cos^2\theta$ factor in the $Z$ mass squared is just the ratio of the
reduced mass of the system to the mass of $0$. In this case, the trick is
just the standard analysis using the reduced mass. It is not clear to me
what the correspondence is for $N>0$.}

Another good check is to derive Georgi-Weinberg
this way. To do this we note that the inverse of $H_\pm$ on the subspace
orthogonal to ${\hat e}$ is
\begin{equation}
H_\pm^{\mbox{\scriptsize``$-1$''}}
= \left(
\left(I-{\hat e}\,{\hat e}^T\right)
- \frac{\left(\kappa-{\hat e}\,(\kappa^T{\hat e})\,\right)\,
\left(\kappa^T-(\kappa^T{\hat e})\,{\hat e}^T\right)}
{(1\pm\kappa^T{\hat e})}\right)
\label{hpmi}
\end{equation}

Now applying this to the $Z$ mass matrix, we use these matrices with
\begin{equation}
\kappa=\kappa_{\np}
\quad\mbox{and}\quad
{\hat e}={\hat e}_{\np}
\label{nps}
\end{equation}
we can transform the neutral gauge boson mass squared matrix to a basis in
which it is orthogonal to ${\hat e}_{\np}$ as follows:\footnote{Remember
from (\protect\ref{toggle}) that $U_\pm^{-1}=U_\pm$.}
\begin{equation}
M_n^2=\frac{1}{4}\,
G\,V\,G\to
U_\pm\,M_n^2\,U_\pm
=\frac{1}{4}\,U_\pm\,
G\,V\,G\,
U_\pm
=
\frac{1}{4}\,H_\pm\,
{\widetilde G}\,\widetilde{\widetilde V}\,{\widetilde G}\,
H_\pm
\label{ztrans}
\end{equation}
The expression (\ref{ztrans}) is interesting for two reasons. Firstly, we
have reduced the $N$$+$$2\times N$$+$$2$
problem to an 
$N$$+$$1\times N$$+$$1$
problem. Secondly, the
expressions are a bit bizarre, with the $(1\pm\kappa^T{\hat e})$. Somehow,
the arbitrary $\pm$ sign must cancel in all physical results.

In particular,
we can trivially invert (\ref{ztrans}) 
on the subspace perpendicular to ${\hat
e}_{\np}$ to get
\begin{equation}
U_\pm\,{M_n^2}^{\mbox{\scriptsize``$-1$''}}\,U_\pm
=
4\,H_\pm^{\mbox{\scriptsize``$-1$''}}\,
{\widetilde G}^{-1}\,{\widetilde{\widetilde
V}}^{\mbox{\scriptsize``$-1$''}}\,{\widetilde G}^{-1}\, 
H_\pm^{\mbox{\scriptsize``$-1$''}}
\label{ztransi}
\end{equation}
Then we can use $U_\pm$ again to transform back to the original basis,
\begin{equation}
4\,U_\pm\,H_\pm^{\mbox{\scriptsize``$-1$''}}\,
{\widetilde G}^{-1}\,{\widetilde{\widetilde
V}}^{\mbox{\scriptsize``$-1$''}}\,{\widetilde G}^{-1}\, 
H_\pm^{\mbox{\scriptsize``$-1$''}}\,U_\pm
\label{mgw3}
\end{equation}
Somewhat miraculously, when one calculates
$H_\pm^{\mbox{\scriptsize``$-1$''}}\,U_\pm$ explicitly, all the $\pm$
dependence cancels and one finds
\begin{equation}
H_\pm^{\mbox{\scriptsize``$-1$''}}\,U_\pm
=
\left(I-{\hat e}\,{\hat e}^T\right)
\,\left(I-\kappa\,\kappa^T\right)
\label{miracle}
\end{equation}
and thus (\ref{mgw3}) reproduces (\ref{mgw1}).

More generally, the expression (\ref{ztransi}) affords a systematic way of
evaluating the inverse mass squared matrix perturbatively. 
The non-zero matrix elements of $H_\pm^{\mbox{\scriptsize``$-1$''}}$ are,
for $j,k=0$ to $N$,
\begin{equation}
\left[H_\pm^{\mbox{\scriptsize``$-1$''}}\right]_{jk}
=\delta_{jk}-\left(\frac{1}{1\pm e/g_{\np}}\right)\,\frac{e^2}{g_jg_k}
\end{equation}
\begin{equation}
\left[{\widetilde G}^{-1}\,H_\pm^{\mbox{\scriptsize``$-1$''}}\right]_{jk}
=\frac{1}{g_j}\delta_{jk}
-\left(\frac{1}{1\pm e/g_{\np}}\right)\,\frac{e^2}{g_j^2g_k}
\end{equation}
Now separating the $0$ components and keeping only the terms that
contribute to second order, we have
\begin{equation}
\left[{\widetilde G}^{-1}\,H_\pm^{\mbox{\scriptsize``$-1$''}}\right]_{00}
=\frac{1}{g_0}
-\left(\frac{1}{1\pm e/g_{\np}}\right)\,\frac{e^2}{g_0^3}
=\frac{1}{g_0}\,\left[1-
\left(\frac{1}{1\pm e/g_{\np}}\right)\,\frac{e^2}{g_0^2}
\right]
\end{equation}
\begin{equation}
\left[{\widetilde G}^{-1}\,H_\pm^{\mbox{\scriptsize``$-1$''}}\right]_{j0}
=
-\left(\frac{1}{1\pm e/g_{\np}}\right)\,\frac{e^2}{g_j^2g_0}
\end{equation}
\begin{equation}
\left[{\widetilde G}^{-1}\,H_\pm^{\mbox{\scriptsize``$-1$''}}\right]_{0k}
=
-\left(\frac{1}{1\pm e/g_{\np}}\right)\,\frac{e^2}{g_0^2g_k}
\end{equation}
\begin{equation}
\left[{\widetilde G}^{-1}\,H_\pm^{\mbox{\scriptsize``$-1$''}}\right]_{jk}
=\frac{1}{g_j}\delta_{jk}
\end{equation}

Expanding
(\ref{ztransi}) to second order in $e/g_j$ for $j=1$ to $N$
and using (\ref{chi}) and (\ref{ttv}), we can collect the
relevant terms of the transformed matrix as follows:
\begin{equation}
\frac{1}{4}\,
\left[U_\pm\,{M_n^2}^{\mbox{\scriptsize``$-1$''}}\,U_\pm\right]_{00}=
\frac{\chi_0}{g_0^2}
\,\left[1-
\left(\frac{1}{1\pm e/g_{\np}}\right)\,\frac{e^2}{g_0^2}
\right]^2
\end{equation}
\begin{equation}
-2\sum_{j=1}^N
\frac{\chi_j}{g_j^2}
\left(\frac{1}{1\pm e/g_{\np}}\right)\,\frac{e^2}{g_0^2}
\,\left[1-
\left(\frac{1}{1\pm e/g_{\np}}\right)\,\frac{e^2}{g_0^2}
\right]
\end{equation}
\begin{equation}
\frac{1}{4}\,
\left[U_\pm\,{M_n^2}^{\mbox{\scriptsize``$-1$''}}\,U_\pm\right]_{0j}=
\left[\frac{\chi_j}{g_0g_j}
-\frac{\chi_0}{g_0g_j}
\left(\frac{1}{1\pm e/g_{\np}}\right)\,\frac{e^2}{g_0^2}
\right]
\,\left[1-
\left(\frac{1}{1\pm e/g_{\np}}\right)\,\frac{e^2}{g_0^2}
\right]
\end{equation}
\begin{equation}
=
\frac{1}{4}\,
\left[U_\pm\,{M_n^2}^{\mbox{\scriptsize``$-1$''}}\,U_\pm\right]_{j0}
\end{equation}
\begin{equation}
\frac{1}{4}\,
\left[U_\pm\,{M_n^2}^{\mbox{\scriptsize``$-1$''}}\,U_\pm\right]_{jk}=
\frac{\chi_{jk}}{g_jg_k}
-\frac{\chi_j+\chi_k}{g_jg_k}
\,\left(\frac{1}{1\pm e/g_{\np}}\right)\,\frac{e^2}{g_0^2}
+\frac{\chi_0}{g_jg_k}
\,\left(\frac{1}{1\pm e/g_{\np}}\right)^2\,\left(\frac{e^2}{g_0^2}\right)^2
\end{equation}

The light $Z$ mass to second order is then given by
\begin{equation}
\frac{1}{4M_Z^2}=
\frac{\chi_0}{g_0^2}
\,\left[1-
\left(\frac{1}{1\pm e/g_{\np}}\right)\,\frac{e^2}{g_0^2}
\right]^2
\end{equation}
\begin{equation}
-2\sum_{j=1}^N
\frac{\chi_j}{g_j^2}
\left(\frac{1}{1\pm e/g_{\np}}\right)\,\frac{e^2}{g_0^2}
\,\left[1-
\left(\frac{1}{1\pm e/g_{\np}}\right)\,\frac{e^2}{g_0^2}
\right]
\end{equation}
\begin{equation}
+\sum_{j=1}^N
\frac{1}{\chi_0}
\left[\frac{\chi_j}{g_j}
-\frac{\chi_0}{g_j}
\left(\frac{1}{1\pm e/g_{\np}}\right)\,\frac{e^2}{g_0^2}
\right]^2
\end{equation}

Expanding and rearranging a bit, this is (note that one of 
the sums now starts at
$j=0$) 
\begin{equation}
\frac{1}{4M_Z^2}
=\frac{1}{\chi_0}\sum_{j=0}^N\frac{\chi_j^2}{g_j^2}
\label{cz-1}
\end{equation}
\begin{equation}
-2
\,\left(\frac{1}{1\pm e/g_{\np}}\right)\,\frac{e^2}{g_0^2}
\,\left[2-
\left(\frac{1}{1\pm e/g_{\np}}\right)\,\frac{e^2}{g_0^2}
\right]
\,\sum_{j=1}^N\frac{\chi_j}{g_j^2}
\label{cz-2}
\end{equation}
\begin{equation}
-\frac{\chi_0}{g_0^2}
\,
\left(\frac{1}{1\pm e/g_{\np}}\right)\,\frac{e^2}{g_0^2}
\,\left[2-
\left(\frac{1}{1\pm e/g_{\np}}\right)\,\left(\frac{e^2}{g_0^2}
+\sum_{j=1}^N \frac{e^2}{g_j^2}\right)\right]
\label{cz-3}
\end{equation}
We can now use the fact that
\begin{equation}
\frac{e^2}{g_0^2}=1-\frac{e^2}{g_{\np}}
-\sum_{j=1}^N\frac{e^2}{g_j^2}
\end{equation}
and substitute this inside the square brackets in (\ref{cz-2}) and
(\ref{cz-3}) to rewrite (\ref{cz-1}-\ref{cz-3}) as
\begin{equation}
\frac{1}{4M_Z^2}
=\frac{1}{\chi_0}\sum_{j=0}^N\frac{\chi_j^2}{g_j^2}
\label{cz2-1}
\end{equation}
\begin{equation}
-2
\,\left(\frac{1}{1\pm e/g_{\np}}\right)\,\frac{e^2}{g_0^2}
\,\left[2-
\left(\frac{1-e^2/g_{\np}^2}{1\pm e/g_{\np}}\right)
\right]
\,\sum_{j=1}^N\frac{\chi_j}{g_j^2}
\label{cz2-2}
\end{equation}
\begin{equation}
-\frac{\chi_0}{g_0^2}
\,
\left(\frac{1}{1\pm e/g_{\np}}\right)\,\frac{e^2}{g_0^2}
\,\left[2-
\left(\frac{1-e^2/g_{\np}^2}{1\pm e/g_{\np}}\right)
\right]
\label{cz2-3}
\end{equation}
where we have neglected higher order terms in (\ref{cz-3}). Now inevitably,
but still apparently miraculously, the $\pm$ signs disappear, and we
have\footnote{Again we have changed the lower limit of a sum.}
\begin{equation}
\frac{1}{4M_Z^2}
=\frac{1}{\chi_0}\sum_{j=0}^N\frac{\chi_j^2}{g_j^2}
-2
\,\frac{e^2}{g_0^2}
\,\sum_{j=0}^N\frac{\chi_j}{g_j^2}
+\frac{\chi_0}{e^2}
\,
\left(\frac{e^2}{g_0^2}\right)^2
\label{cz3}
\end{equation}

To the order to which we are working, we can replace the factors of
$e^2/g_0^2$ in (\ref{cz3}) with any expression that has the same zeroth
order value. It is convenient to substitute
\begin{equation}
\frac{e^2}{g_0^2}\to
\frac{e^2}{\chi_0}\,\sum_{j=0}^N\frac{\chi_j}{g_j^2}
\end{equation}
which simplifies (\ref{cz3}) further to
\begin{equation}
\frac{1}{4M_Z^2}
=\frac{1}{\chi_0}\sum_{j=0}^N\frac{\chi_j^2}{g_j^2}
-
\frac{e^2}{\chi_0}
\,\left(\sum_{j=0}^N\frac{\chi_j}{g_j^2}\right)^2
\label{cz4}
\end{equation}

Note also that the inverse mass matrix for the heavy neutral gauge bosons
to leading order is an $N\times N$ matrix with matrix elements
\begin{equation}
4\left(\frac{\chi_{jk}}{g_jg_k}
-\frac{\chi_j+\chi_k}{g_jg_k}
\,\left(\frac{1}{1\pm e/g_{\np}}\right)\,\frac{e^2}{g_0^2}
+\frac{\chi_0}{g_jg_k}
\,\left(\frac{1}{1\pm e/g_{\np}}\right)^2\,\left(\frac{e^2}{g_0^2}\right)^2
\right.
\end{equation}
\begin{equation}
\left.
-\frac{1}{\chi_0}\,
\left[\frac{\chi_j}{g_j}
-\frac{\chi_0}{g_j}
\left(\frac{1}{1\pm e/g_{\np}}\right)\,\frac{e^2}{g_0^2}
\right]\,
\left[\frac{\chi_k}{g_k}
-\frac{\chi_0}{g_k}
\left(\frac{1}{1\pm e/g_{\np}}\right)\,\frac{e^2}{g_0^2}
\right]\right)
\end{equation}
\begin{equation}
=4\,\frac{1}{g_j}\,\left(
\chi_{jk}-\frac{\chi_j\,\chi_k}{\chi_0}
\right)\,\frac{1}{g_k}
\label{heavy-z-matrix}
\end{equation}
for $j,k=1$ to $N$. This looks just like (\ref{heavy-z-matrix}), as it
should to this order.

\setcounter{equation}{0}\section{\label{alphabet}Alphabet soup}

Let's review where we are. From the low energy 
charged-current weak interactions, we have (pulling some equations from
previous sections with their original numbers)
$$
\sqrt{2}\, G_F=\frac{1}{v^2}
=[\widetilde V^{-1}]_{00}
\leadsto\sum_{j=0}^{N}\frac{1}{v_{j,j+1}^2}
\eqno{(\protect\ref{vi})}
$$
The low energy neutral-current weak interactions are
$$
\sum_{j,k=0}^{N}[\widetilde V^{-1}]_{jk}
\left[T_3\delta_{j0}-\frac{e^2}{g_j^2}Q\right]\,
\left[T_3\delta_{k0}-\frac{e^2}{g_k^2}Q\right]
\eqno{(\protect\ref{nc})}
$$
where the matrix elements $[\widetilde V^{-1}]_{jk}$ are given by
$$
[\widetilde V^{-1}]_{00}=\frac{1}{v^2}\equiv \chi_0\,,
\quad
[\widetilde V^{-1}]_{j0}=
[\widetilde V^{-1}]_{0j}\equiv \chi_j\,,
\quad
[\widetilde V^{-1}]_{jk}\equiv \chi_{jk}
\quad\mbox{for $j,k=1$ to $N$}
\eqno{(\protect\ref{chi-gen})}
$$
Note that the normalization of the $T_3^2$ term satisfies custodial
$SU(2)$ symmetry, so the correction to the $\rho$ 
parameter is small. The analog of
$\sin^2\theta$ as determined by the low energy weak interactions is
determined by the coefficient of $T_3\, Q$ in (\ref{nc}) to be
\begin{equation}
\sin^2\theta=e^2v^2\sum_{j=0}^{N}\frac{\chi_j}{g_j^2}
\label{sin2theta}
\end{equation}
The $W$ and $Z$ masses are determined to next-to-leading order in the small
couplings by
$$
\frac{1}{M_W^2}=4v^2\,\sum_{j=0}^{N}\frac{\chi_j^2}{g_j^2}
\eqno{(\protect\ref{mw2i})}
$$
\begin{equation}
\frac{1}{M_Z^2}
=4v^2\left(
\sum_{j=0}^N\frac{\chi_j^2}{g_j^2}
-
\frac{e^2}{\chi_0}
\,\left(\sum_{j=0}^N\frac{\chi_j}{g_j^2}\right)^2
\right)
\label{mz2}
\end{equation}

One way of describing this is to say that we can write all quantities to this
order in terms of four parameters (as usual, shown in general and with
their values in the linear model):
\begin{eqnarray}
&&e^2=\left(\sum_{k=0}^{N+1}\frac{1}{g_k^2}\right)^{-1}
\label{param-e}
\\
&&v^2=\frac{1}{\chi_0}\leadsto
\left(\sum_{k=0}^{N}\frac{1}{v_{k,k+1}^2}\right)^{-1}
\label{param-v}
\\
&& s_1^2\equiv e^2v^2\sum_{j=0}^{N}\frac{\chi_j}{g_j^2}
\leadsto\frac{e^2}{g_0^2}+
\sum_{j=1}^{N}\frac{e^2}{g_j^2}
\left(
\sum_{k=j}^{N}\frac{v^2}{v_{k,k+1}^2}
\right)
\label{param-s1}
\\
&& s_2^2\equiv e^2v^4\sum_{j=0}^{N}\frac{\chi_j^2}{g_j^2}
\leadsto\frac{e^2}{g_0^2}+
\sum_{j=1}^{N}\frac{e^2}{g_j^2}
\left(
\sum_{k=j}^{N}\frac{v^2}{v_{k,k+1}^2}
\right)^2
\label{param-s2}
\\
\end{eqnarray}
The two parameters, $s_1^2$ and $s_2^2$ both reduce to
$\sin^2\theta$ in the standard model limit in which the extra gauge
couplings go to infinity. In terms of these, we can write
\begin{equation}
\sin^2\theta_{\rm neutral\atop current}=s_1^2
\label{ncs2}
\end{equation}
\begin{equation}
M_W^2=\frac{e^2v^2}{4s_2^2}
\label{mws1}
\end{equation} 
\begin{equation}
M_Z^2=
\frac{e^2v^2}{4(s_2^2-s_1^4)}
=\frac{s_2^2}{s_2^2-s_1^4}M_W^2
\label{mzs123}
\end{equation}

Note that $s_1^2>s_2^2$.

According to the particle data group,~\cite{PDBook}
\begin{equation}
\begin{array}{l}
\displaystyle
M_Z^2=M_{Z0}^2\frac{1-\alpha T}{1-G_Fm_{Z0}^2S/2\sqrt{2}\pi}
\\ \displaystyle
M_W^2=M_{W0}^2\frac{1}{1-G_Fm_{W0}^2(S+U)/2\sqrt{2}\pi}
\end{array}
\end{equation}
``where $M_{Z0}$ and $M_{W0}$ are the SM expressions (as functions of $m_t$
and $M_H$) in the {\small$\overline{MS}$} scheme.''
Or more sloppily (but simpler for our purposes)
\begin{equation}
\begin{array}{l}
\displaystyle
M_Z^2=\frac{(1-\alpha T)e^2v^2/4}
{\sin^2\theta\cos^2\theta-\alpha S/4}
\\ \displaystyle
M_W^2=\frac{e^2v^2/4}
{\sin^2\theta-\alpha (S+U)/4}
\end{array}
\label{stu}
\end{equation}

All this implies the following.
\begin{enumerate}
\item The
custodial $SU(2)$ relation for the ratio of NC to CC low energy weak
interactions, along with the fact (see the discussion of (\ref{gw}) below)
that the heavy gauge bosons make a negligible contribution
implies that to this order,
\begin{equation}
T=0
\label{t}
\end{equation}
\item 
From the expressions for $M_W^2$ and $\sin^2\theta$, (\ref{mzs123}) and
(\ref{ncs2}), we get $S$,\footnote{This formula was derived in the special
case of the linear model in \cite{Hirn:2004ze}.}
\begin{equation}
s_2^2=s_1^2-\alpha S/4
\label{s1}
\end{equation}
or
\begin{equation}
S=\frac{4}{\alpha}(s_1^2-s_2^2)
\label{s}
\end{equation}
In particular this shows that $S>0$. This result is well known for the
linear model.~\cite{Foadi:2003xa, Chivukula:2004pk} We now see that it is
true for more general symmetry breaking patterns, at least near the
standard model limit. 
\item
Now from the expression from $M_Z^2$ and the rest, we can write
\begin{equation}
s_2^2-s_1^4
=s_1^2-s_1^4-\alpha (S+U)/4
\label{u1}
\end{equation}
or
\begin{equation}
U=0
\label{u2}
\end{equation}
The right hand side of (\ref{u2}) is higher order in the small quantities
$e^2/g_j^2$ for $j=1$ to $N$.
\end{enumerate}

Comparing (\ref{s}), (\ref{param-s1}), (\ref{param-s2}) and (\ref{hw}), we
can write
\begin{equation}
S-4\pi^2v^2\sum_{{\rm heavy}\atop W{\rm s} }\frac{1}{M^2}
=16\pi^2
v^2\left(
\sum_{j=0}^{N}\frac{\chi_j}{g_j^2}
-
\sum_{j=0}^{N}\frac{\chi_{jj}}{g_j^2}
\right)\leadsto0
\end{equation}
Thus the $S$ parameter is related to the sum of the inverse mass
squares of the heavy gauge
bosons only in the linear model. In general these are independent.

The couplings of the $W$ and $Z$ to fermions are determined by
the mass eigenstates. We know the $W$ eigenstate from 
(\ref{weigenvector}).
We find for the coupling
\begin{equation}
g_W^2=\chi_0^2\Biggm/
\sum_{j=0}^{N}\frac{\chi_j^2}{g_j^2}
=\frac{M_W^2}{v^2}
\label{gw}
\end{equation} 
which is the same as the tree-level standard model result.
This does not get corrections to this order because the heavier states
are not only heavier, they are more weakly coupled to ordinary matter.
Presumably, the $Z$ couplings behave the same way for the same reason. 
This is more complicated
to work out, so I won't do it here. 

\setcounter{equation}{0}\section{\label{mechanicals}The mechanical analog
of $S$} 

How small can we make $S$? To think about this, let us first rewrite
(\ref{s}) as
\begin{equation}
S=\frac{1}{\pi}\sum_{j=1}^{N}\left(\frac{(4\pi)^2}{g_j^2}\right)
\,\left(\frac{\chi_j}{\chi_0}\right)
\,\left(1-\frac{\chi_j}{\chi_0}\right)
\label{ss}
\end{equation}
Even if we take the coupling factors to be of order $1$, the terms in the
sum are each of order $1/4$ unless the $\chi_j/\chi_0$ is close to $0$ or
$1$, in which case the contribution is small.\footnote{A similar formula is
written down in \cite{Casalbuoni:2004id}. I would like to thank the authors
for their comments about this.}

In the linear model with approximately equal $v_{j,j+1}^2$ along the line,
each $v_{j,j+1}^2$ 
is approximately $(N+1)v^2$, so the Higgs mass can be raised by a
factor of $\sqrt{N+1}$, and
\begin{equation}
\frac{\chi_j}{\chi_0}\approx\frac{N+1-j}{N+1}
\end{equation}
so that
\begin{equation}
S\gtrsim\frac{N^2-1}{6\pi N}
\end{equation}
which suggests that $N$ cannot be large.

It might not be immediately obvious that
one cannot do significantly better than this 
in a completely general model. But in fact, this becomes quite clear if
you think about what this means in the mechanical analog. The $\chi_j$ are
the components of the low-frequency normal mode of the mechanical analog,
so $\chi_j/\chi_0$ is the displacement of the $j$th mass as a fraction of
the displacement of the $0$th mass when the system is stretched slowly by
pulling on the $0$th mass. Thus to get small $S$, we want a system in which
all the masses 
either move very little when the $0$th mass is pulled, or else move
along with the $0$th mass. What we don't want is a number of masses whose
motions interpolate between the motion of mass $0$ and the fixed wall,
because these give the maximum contribution to (\ref{ss}). Now it is perfectly
possible to have a system of
springs with the properties that give small $S$
(for example, the systems analogous to
figures~\ref{fig-1-3}a and \ref{fig-1-3}b, where the spring to mass $1$
would not be stretched at all), but unfortunately it is not
consistent with 
the fundamental goal of producing a low frequency mode with only stiff
springs (that is - raising the Higgs mass). The low frequency mode arises
precisely because the stretching of the system can be spread over many
stiff springs, so that each stretches only a small amount. But that means
that the displacements vary from zero to the full displacement. 

Thus I conclude that
raising the Higgs mass is essentially equivalent to increasing $S$, not
just in the linear system, or in any other deconstruction of one or more
extra dimensions, but for the completely general structure of $SU(2)$s and
a single $U(1)$. And while we have had some fun with Higgsless theories, it
seems unlikely that nature has chosen this amusing approach to electroweak
symmetry breaking.

\section*{Acknowledgements} I am grateful for comments from Nima
Arkani-Hamed, Sekhar Chivukula, Elizabeth Simmons and Devin Walker. 
This research is supported in part by
the National Science Foundation under grant PHY-0244821.

\bibliography{higgsless}

\end{document}